\documentclass[iop,jcp,twocolumn,floatfix]{revtex4-1} 
\usepackage[dvipdf]{graphicx}
\usepackage{multirow}
\usepackage{array}
\usepackage[usenames]{color}
\usepackage{amsmath}
\usepackage{amssymb}
\usepackage{ulem}
\newcommand{\avg}[1]{\ensuremath{\left<#1\right>}}

\begin{document}
\newlength{\imgsz}
\setlength{\imgsz}{0.9\columnwidth}

\title{Efficient methods and practical guidelines for simulating isotope effects}

\author{Michele Ceriotti}
\email{michele.ceriotti@chem.ox.ac.uk}
\affiliation{Physical and Theoretical Chemistry Laboratory, 
University of Oxford, South Parks Road, Oxford OX1 3QZ, UK}

\author{Thomas E. Markland}
\email{tmarkland@stanford.edu}
\affiliation{Department of Chemistry, Stanford University, Stanford, CA 94305, USA}

\begin{abstract}
The shift in chemical equilibria due to isotope substitution is often exploited 
to gain insight into a wide variety of chemical and physical processes. 
It is a purely quantum mechanical effect, which can be computed exactly using simulations
based on the path integral formalism. Here we discuss how these techniques can be made dramatically more efficient, 
and how they ultimately outperform quasi-harmonic approximations to treat quantum 
liquids not only in terms of accuracy, but also in terms of computational efficiency.
To achieve this goal we introduce path integral quantum mechanics estimators based 
on free energy perturbation, which enable the evaluation of isotope effects 
using only a single path integral molecular dynamics trajectory of the naturally abundant isotope.
We use as an example the calculation of the free energy change 
associated with H/D and $^{16}$O/$^{18}$O substitutions in liquid water, and of the 
fractionation of those isotopes between the liquid and the vapor phase.
In doing so, we demonstrate and discuss quantitatively the relative benefits of each approach,
thereby providing a set of guidelines that should facilitate the choice of the most appropriate
method in different, commonly encountered scenarios.
The efficiency of the estimators we introduce and the analysis that we perform
should in particular facilitate accurate {\it ab initio} calculation of isotope effects 
in condensed phase systems.
\end{abstract}

\maketitle

\section{Introduction}

Replacing an element with one of its isotopes is one of the most useful 
tools to study chemical mechanisms and equilibria.
Since isotopes differ only by the number of neutrons, the potential energy surface on which they evolve, 
which arises from the electron-proton interactions, is unchanged. 
If the nuclei behaved as classical particles, moving on the Born-Oppenheimer
potential energy surface, the thermodynamic properties of systems containing different isotopes
would be identical. However, light nuclei exhibit nuclear quantum effects such
as zero-point energy and tunnelling, which are more pronounced for lighter isotopes. 
As such, isotope effects provide an experimental window 
through which the influence of nuclear quantum effects can be probed.

Accurate and efficient simulations of isotope effects would enable the investigation 
of many important processes such as acid-base chemistry, shifts in phase boundaries, 
geological fractionation ratios and equilibrium constants of chemical reactions \cite{Chacko2001,Wolfsberg2009}. 
In addition, the shift in the free energy barrier along a particular reaction 
coordinate can be used to obtain the change in the rate of a chemical reaction via 
a quantum transition state theory \cite{Gillan1987,Voth1989,Mills1997} or, when combined with a dynamics scheme such 
as centroid molecular dynamics or ring polymer molecular dynamics, an approximation 
to the full dynamical rate constant \cite{cao-voth94jcp,crai-mano04jcp}. Isotope effects in these processes 
show a range of interesting behaviors. For example recent work on water and other hydrogen-bonded 
systems have demonstrated that quantum mechanical effects can act to strengthen short hydrogen bonds 
or to weaken long ones \cite{Chen2003,habe+09jcp,li+11pnas}. 
This concept of ``competing quantum effects'' has recently been 
shown to explain the inversion in the atmospherically important fractionation of hydrogen and 
deuterium between the liquid and vapor phases of water where at ambient temperatures D is favored 
in the liquid whereas at higher temperatures D is favored in the vapor \cite{mark-bern12pnas}. Accurately 
describing the cancellation between the quantum effects in water
 requires the inclusion of anharmonic terms in the potential and  
therefore conclusions drawn from simple harmonic approximations can be misleading. In addition the 
subtle nature of this cancellation means that fractionation ratios can be used as a sensitive 
test of a potential energy surface's ability to accurately predict quantum effects. 

Modelling isotope effects requires the calculation of the free energy change upon isotope 
substitution. For some systems such as solids and simple molecular gases a reasonable approximation 
of the isotope effect can be gained from either a harmonic approximation or $\hbar$ perturbation \cite{Wigner1932,Kirkwood1933,Herzfeld1938}. 
However for liquids, which contain many anharmonic modes, or when treating 
more strongly quantum mechanical atoms such as H and D, these approximations can be inaccurate 
or inefficient \cite{Chialvo2009,mark-bern12pnas}. 

The exact calculation of isotope exchange free energies can be 
achieved by using the imaginary time path integral formalism of quantum 
mechanics \cite{feyn-hibb65book,chan-woly81jcp,parr-rahm84jcp}. This formalism 
exploits the isomorphism between a quantum mechanical 
system and a classical system in an extended ``ring polymer'' phase space. As such it 
can be applied to any system for which a classical simulation can be performed, albeit 
traditionally with a significant computational overhead.

A series of recent developments have dramatically decreased the computational cost of 
path integral simulations. These have taken the form of the introduction of efficient 
integrators and thermostatting schemes to address 
non-ergodicity \cite{tuck+93jcp,mart+99jcp,ceri+10jcp}. In addition, 
methods have been developed to reduce the number of force evaluations or points on the 
imaginary time path \cite{Markland2008,Markland2008b,Fanourgakis2009,ceri+09prl2,ceri+11jcp,ceri-mano12prl}. 
These developments along with improvements in the ability to 
perform computationally efficient {\it ab initio} calculations make on-the-fly first principles path 
integral molecular dynamics a feasible proposition \cite{Marx1996,Marx1999,Ufimtsev2008,Fujita2009}. 
Such a combination is natural, 
since {\it ab initio} calculations attempt to reproduce the exact Born-Oppenheimer potential energy surface.
Whereas nuclear quantum effects are often implicitly accounted for by empirical potential energy surfaces,
which are typically parameterized to reproduce experimental properties, they are completely
absent in {\it ab initio} calculations unless they are treated explicitly.

In this work we introduce a series of developments to allow the extraction of free energy changes upon 
isotope substitution from path integral simulations. Firstly, we show how the 
free energy change can be computed efficiently by performing a physically 
motivated transformation of the thermodynamic integration variables. 
Secondly, we address the failures of harmonic approximations when applied to liquids and other 
systems with anharmonic degrees of freedom in terms of both accuracy and, more remarkably, 
efficiency when compared to path integral simulations. 
Thirdly, we introduce two path integral estimators, based on a free energy 
perturbation approach, that offer desirable features such as the ability to extract isotope free 
energy changes from a single path integral trajectory of the most abundant isotope.
We benchmark these methods against the previously introduced thermodynamic integration
 approach \cite{Vanicek2007} using H/D substitution in liquid water as a prototypical model of a 
hydrogen bonded system. In doing so, we demonstrate the strengths and pitfalls which each 
possess, and identify which approach may be most advantageous for a given application. 

\section{Theory}
\label{sec:theory}

Let us consider the thermodynamics of substituting an atom $X$ of mass $m$ in a system $\Phi$ with a different 
isotope of the same element, having mass $m'$. More specifically, we will consider the process of taking an atom in a phase, molecule or compound, which we will label as $^m\text{X}(\Phi)$, and exchanging it with an atom of the other isotope taken from a reservoir of non-interacting atoms,
which we will label  $^{m'}\text{X}(\infty)$:
\begin{equation}
^{m}\text{X}(\Phi) + ^{m'}\text{X}(\infty) \rightleftharpoons ^{m'}\text{X}(\Phi) + ^{m}\text{X}(\infty)
\end{equation}
The equilibrium is controlled by the free energy change corresponding 
to this process, $\Delta A$. Here we will consider the transformation to be performed
at constant volume, but the extension to constant pressure is straightforward. 
One can compute $\Delta A$ by thermodynamic integration, by first 
writing out the quantum mechanical free energy at inverse temperature $\beta$ 
for the system $\Phi$ containing an isotope of mass $\mu$
\begin{equation}
A(\mu,\Phi)=-\frac{1}{\beta}\ln \mathrm{Tr} \left[ e^{-\beta H(\mu,\Phi)} \right],
\end{equation} 
and then differentiating with respect to mass, which yields an expression
containing the expectation value of the quantum kinetic energy of the 
system:
\begin{equation}
\frac{\partial A(\mu,\Phi)}{\partial\mu}=-\frac{\avg{T(\mu,\Phi)}}{\mu}.
\end{equation} 
The thermodynamic integration for the free atoms is straightforward, and one is left 
with an expression for the free energy difference that reads
\begin{equation}
\Delta A(\Phi)=\frac{d}{2\beta } \log\frac{m'}{m}-\int_m^{m'} \frac{\avg{T(\mu,\Phi)}}{\mu} \mathrm{d}\mu.
\label{eq:iso-a}
\end{equation}
Here $d$ is the dimensionality of the problem and $\avg{T(\mu,\Phi)}$ is the
expectation value of the kinetic energy of atom $X$ with mass $\mu$ in the system $\Phi$.

One observation that is immediately clear from this result is that if the atom behaved classically, its 
average kinetic energy would yield the classical equipartition value of $1/2\beta$ per 
degree of freedom and the free energy difference in Eq.~\eqref{eq:iso-a} would amount to zero. 
In order to compute $\Delta A$ it is therefore necessary 
to evaluate the expectation value of the kinetic energy in a way that takes into account the
quantum mechanical nature of the atoms.

\subsection{Discretizing the free energy change} \label{sub:th-int}

One can in practice discretize the integral in Eq.~\eqref{eq:iso-a} and compute the integrand for a few values of the isotope mass, $\mu$. 
In previous studies a linear discretization in the isotope mass $\mu$ has been 
used \cite{Vanicek2007,Paesani2009,Perez2011,mark-bern12pnas}. 
However, one can accelerate the convergence considerably by 
performing a change of variables that takes into account the physical 
nature of the problem. Making such a change of variables acts to smooth 
the integrand so that a good approximation can be obtained with a very small number of integration points. 

\begin{figure}
\caption{\label{fig:th-int} Panel (a) shows the integrand of Eq.~\eqref{eq:iso-a}, 
as computed for a $P=64$ PIMD simulation of liquid water with one hydrogen atom
being transformed into deuterium. Panel (b) shows that the integrand becomes nearly
constant when a transformed integration variable is used, as in Eq.~\eqref{eq:ti-transformed}.
Lines are just guides for the eye, and statistical error bars are reported for each point.
 The dashed line shows the estimate from the two end points. 
Note the difference in the scale: in panel (a) the integrand varies by a factor of 2, 
whereas in panel (b) it varies by less than 3\%. }
\includegraphics[width=1.0\columnwidth]{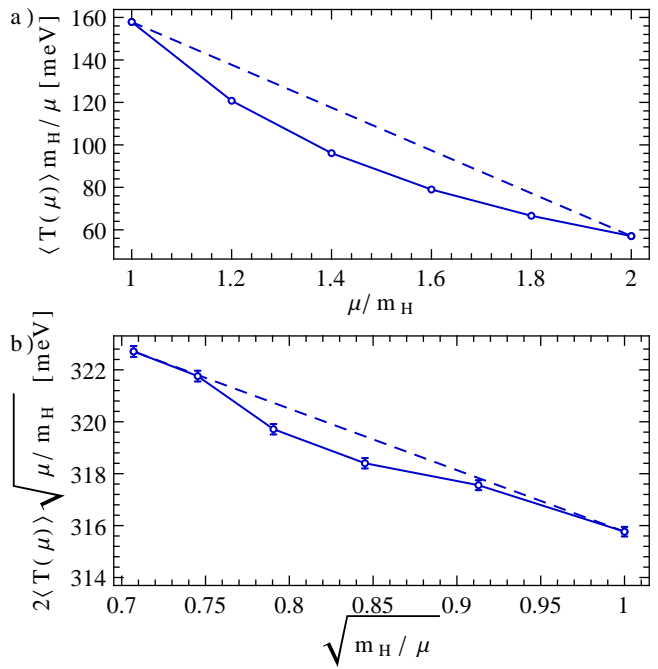}
\end{figure}

If the mass-dependence of $\avg{T(\mu)}$ were known, one could make the integrand
$\avg{T(\mu)}/\mu$ constant by performing the transformation $y=f^{-1}(\mu)$ where 
$f(y)$ satisfies the differential equation
\begin{equation}
\avg{T\left(f(y)\right)} f'(y)/f(y) = C,
\label{eq:ti-sde}
\end{equation}
where C is any constant. Clearly $\avg{T(\mu)}$ is not known exactly beforehand, so one cannot in practice solve Eq.~\eqref{eq:ti-sde}. 
However, one can use physical intuition to make assumptions on the form of $\avg{T(\mu)}$ and derive a transformation
which ``flattens'' the integral. A particularly useful limit for condensed phase systems is the limit where the problem  
is nearly harmonic and strongly quantized. In that case it can be shown that
\begin{equation}
\avg{T(\mu)}\propto \sum_i \sqrt{k_i/\mu},
\end{equation}
where the $k_i$'s are the spring constants of the various normal modes. With this assumption
one obtains $f(y)\propto {4(\sum_i \sqrt{k_i})^2}/{(C y+B \sum_i \sqrt{k_i})^2}$, where $B$ is an integration constant. 
In practice, by choosing $B=0$ and $C=2\sum_i \sqrt{k_i}$, the desired change of
variable is simply $\mu=1/y^2$. Eq.~\eqref{eq:iso-a} is therefore transformed into
\begin{equation}
\Delta A=\frac{d}{2\beta } \log\frac{m'}{m}-\int_{1/\sqrt{m'}}^{1/\sqrt{m}} 2\frac{\avg{T(1/y^2)}}{y} \mathrm{d}y.
\label{eq:ti-transformed}
\end{equation} 
Figure~\ref{fig:th-int} shows the integrand of Eqs.~\eqref{eq:iso-a} and~\eqref{eq:ti-transformed}
for the H/D substitution in liquid water.  
Using the coordinate transformation in Eq.~\eqref{eq:ti-transformed} makes the integrand nearly constant, 
which simplifies greatly the integration. Using Eq.~\eqref{eq:iso-a}, one would obtain $\Delta A=-94.0\pm0.1$~meV 
when computing six integration points, and $-107.5\pm0.1$~meV where only the two extrema are used. 
From Eq.~\eqref{eq:ti-transformed} one would instead obtain the fully converged value $\Delta A=-93.4\pm 0.1$~meV 
when using six points, and $-93.5\pm 0.1$~meV when using only two.

\subsection{The pitfalls of harmonic approximations}\label{sub:harm}

For solids and gases nuclear quantum effects are usually accounted for using
the harmonic approximation \cite{pavo-baro94ssc}. Within this approximation, the quantum kinetic energy can
be estimated by first computing the normal mode frequencies $\omega_i$ and the 
corresponding normalized eigenvectors $\mathbf{u}^{(i)}$ and then evaluating the kinetic energy for a given  
atom $k$ using \cite{ceri+10prb,lin+11prb,krzy-fern11prb,seel+12jpcm}
\begin{equation}
T_{k}^{\text{harm}} = \sum_{i\alpha} \left|u^{(i)}_{k\alpha}\right|^2 \frac{\hbar\omega_i}{4}\coth 
\frac{\beta\hbar\omega_i}{2}.
\label{eq:t-harm}
\end{equation}
Here $u^{(i)}_{k\alpha}$ indicates the component of the eigenvector $\mathbf{u}^{(i)}$ corresponding to atom
$k$ in the Cartesian direction $\alpha$. In solids, one would need to integrate over the Brillouin zone. 
For gas-phase molecules one can include the (classical) contributions from rotations and translations by
also summing over the zero-frequency eigenvalues of the dynamical matrix. 
Including quantized rotations adds considerable additional complexity but is only necessary at very low temperatures.

\begin{figure}
\caption{\label{fig:harmonic} The figure shows the integrand of the transformed
thermodynamic integration~\eqref{eq:ti-transformed} as a function of the mass of the isotope.
The results from fully-converged PIMD ($P=64$, blue lines) are compared with those from 
a harmonic/quasi-harmonic approximation (red lines). Curves are shown for both liquid water (full lines)
and an isolated molecule in the gas phase (dashed lines). The harmonic approximation for the gas phase
is based on static calculations and therefore has no error bars. For all other calculations lines are just guides for 
the eye, and statistical error bars are reported for each data point.  }
\includegraphics[width=1.0\columnwidth]{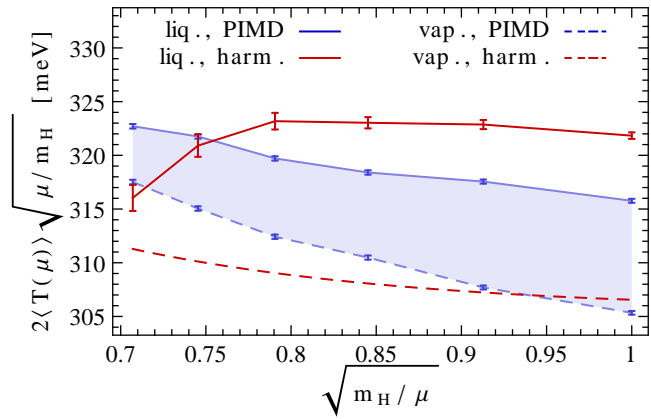}
\end{figure}

For solids and simple gas-phase molecules this is often a good approximation which can be computed inexpensively.
However, the cost of computing the Hessian grows rapidly with the number of atoms. 
Moreover, in systems such as liquids that contain many anharmonic modes it makes little sense to compute the Hessian 
for finite-temperature configurations \cite{poho+87jcp}. 
However, one can compute the vibrational density of states for a given 
atom from the Fourier transform of the velocity-velocity correlation function, 
\begin{equation}
D_k(\omega)\propto 
\int_{-\infty}^{\infty}\left<\mathbf{v}_k(t)\cdot\mathbf{v}_k(0)\right> e^{-\mathrm{i}\omega t} \mathrm{d}t 
\end{equation}
where $\mathbf{v}_k$ is the velocity of atom $k$. 
An approximation to the quantum kinetic energy can then be computed by normalizing $D_k(\omega)$ 
such that  $\int_0^\infty D_k(\omega) \mathrm{d}\omega =3$ and evaluating \cite{krzy-fern11prb}
\begin{equation}
T_{k}^{\text{DOS}} = \int_0^\infty D_k(\omega) \frac{\hbar\omega}{4}\coth \frac{\beta\hbar\omega}{2} \mathrm{d}\omega.
\label{eq:t-dos}
\end{equation}
Figure~\ref{fig:harmonic} contrasts the values of the integrand that enter  Eq.~\eqref{eq:ti-transformed}
for both liquid water and water vapor at $300$~K, as computed exactly by direct substitution (see Sec.~\ref{sub:direct}) 
and by quasi-harmonic analysis (Eq.~\eqref{eq:t-harm} for the vapor and Eq.~\eqref{eq:t-dos} for the liquid). 
For each phase, the harmonic approximation is close to the corresponding exact path integral result.
There are however two important observations to be made.

Firstly, one is generally interested in computing \emph{differences} between isotope-substitution free energies 
such as isotopic fractionation ratios or acidity shifts, which are highly sensitive to subtle anharmonic effects. 
For instance the geochemically important process of isotopic fractionation between liquid water and its 
vapor corresponds to the free energy change associated with the process
\begin{equation}
\text{H}_{2}\text{O}(l) + \text{HOD}(v) \rightleftharpoons \text{H}_{2}\text{O}(v) + \text{HOD}(l)
\label{eq:frac-wat}
\end{equation}
where $l$ denotes the liquid phase, and $v$ denotes the vapor phase. 
Typically this is expressed as a fractionation ratio,
\begin{equation}
\alpha_{l-v} = e^{-\beta\left(\Delta A(l)-\Delta A(v)\right)}.
\label{eq:frac-fac}
\end{equation}
Using the data in Fig.~\ref{fig:harmonic} to compute the fractionation ratio the quasi-harmonic 
approximation yields a value of $10^3\log \alpha_{l-v}=156\pm10$ compared to the exact path integral 
result of $95\pm 1$. A harmonic approximation would  therefore substantially overpredict the amount 
to which H and D separate in the atmosphere.

Secondly, although one might view the quasi-harmonic approximation in Eq.~\eqref{eq:t-dos} 
as a computationally inexpensive approach to obtain an estimate of the quantum kinetic energy of 
a liquid, this is not the case. The reason for its inefficiency is that the kinetic energy 
is obtained from the integral of the Fourier transform of the velocity autocorrelation 
function. Hence any statistical noise in the velocity autocorrelation function must be 
very small to avoid the appearance of large statistical errors in the quantum kinetic energy. 
For example to 
obtain the exact results in Fig.~\ref{fig:harmonic} we performed a $P=64$ bead path 
integral trajectory which in the most naive implementation is $\sim64$ times more costly 
than a single classical simulation of the same length. Despite the fact that we computed
the quasi-harmonic estimate in Eq.~\eqref{eq:t-dos} from 64 independent classical trajectories 
of the same length -- which thereby required the same computational effort as the path 
integral simulation -- the statistical error of the PIMD results is 
10 times smaller than that of the quasi-harmonic approximation. 
In other terms, for a given level of statistical accuracy, a quasi-harmonic approximation
such as in Eq.~\eqref{eq:t-dos} is 100 times more demanding than a fully converged PIMD
simulation, even if we considered a number of beads twice as large as what is generally
deemed converged and we did not exploit the efficient estimators we introduce in Section~\ref{sub:fep-estimators},
ring-polymer contraction \cite{mark-mano08cpl} or the PIGLET method \cite{ceri-mano12prl}.

In summary, while harmonic approximations may be a viable way to crudely estimate the
role of nuclear quantum effects and the change in isotope substitution free energies in the gas 
phase and in simple, harmonic solids by means of Eq.~\eqref{eq:t-harm} they should be treated 
with caution in terms of both efficiency and accuracy for liquids and systems with anharmonicity. 
In particular whenever something as subtle
as isotope fractionation ratios are required or whenever the system is strongly anharmonic, 
the path integral methods we discuss in the following section are not only more accurate, but paradoxically
less computationally demanding than a quasi-harmonic approximation.

\subsection{Imaginary time path integrals}

To move beyond the harmonic approximation to the kinetic energy in 
Eq.~\eqref{eq:iso-a} while including the quantum nature of the nuclei one can use the 
imaginary time path integral formulation 
of quantum mechanics. This approach allows one to exactly include the quantum nature of the 
nuclei for a system of $N$ distinguishable particles at a finite temperature. 
 For simplicity we shall use a notation for a single particle of mass $m$ in one dimension, 
 described by its position $q$ and momentum $p$. The generalization to multiple degrees 
 of freedom is routine and has been widely discussed elsewhere \cite{parr-rahm84jcp,bern-thir86arpc}. 
 For a system with a Hamiltonian of the form $H(p,q)=p^2/2m + V(q)$, one can show that the quantum mechanical
partition function $Z=\mathrm{Tr} \left[ e^{-\beta H} \right]$ at inverse temperature $\beta$
is isomorphic to the classical partition function
of an extended system -- the so-called \emph{ring polymer}:
\begin{equation}
Z_P\propto \int \mathrm{d}\mathbf{p}\,\mathrm{d}\mathbf{q}\, e^{-\frac{\beta}{P} H_P\left(\mathbf{p},\mathbf{q}\right) }.
\label{eq:pi-z}
\end{equation}
Here $H_P$ is the ring polymer Hamiltonian
\begin{equation}
H_P\left(\mathbf{p},\mathbf{q}\right)=\sum_{i=0}^{P-1} \frac{p_i^2}{2m}+V(q_i)+ 
\frac{1}{2}m\omega_P^2\left(q_i-q_{i+1}\right)^2,
\label{eq:pi-h}
\end{equation}
where cyclic boundary conditions $i+P\equiv i$ are implied, and $\omega_P=P/\beta\hbar$.
Equation~\eqref{eq:pi-h} describes $P$ replicas (beads) of the physical system, where replicas with adjacent
indices are connected by harmonic springs to form a closed loop. As $P\rightarrow \infty$ the 
path integral partition function $Z_P$ converges to the correct quantum mechanical one.

The momenta in the partition function of Eq.~\eqref{eq:pi-z} simply provide a sampling tool and can 
be integrated out analytically. Hence all the quantum mechanical information is encoded in the 
configurational part of $Z_P$. 
Therefore, even a property which is directly related to the momentum such as 
the kinetic energy must be computed by means of a configurational
average of an appropriate estimator. In what follows we will use $\avg{O}_m$ to indicate the 
configurational average of the estimator $O(\mathbf{q})$ over the canonical ensemble for a particle of mass $m$,
\begin{equation}
\avg{O}_m = \frac{\int\mathrm{d}\mathbf{q}\, O(\mathbf{q})\, e^{-\frac{\beta}{P} 
\sum_i V\left(q_i\right) +m\omega_P^2\left(q_i-q_{i+1}\right)^2/2  }  } 
{\int\mathrm{d}\mathbf{q}\, e^{-\frac{\beta}{P} \sum_i V\left(q_i\right) +m\omega_P^2\left(q_i-q_{i+1}\right)^2/2  }}.
\label{eq:avgm}
\end{equation}
This phase-space average can be evaluated by generating a sequence of configurations
consistent with the appropriate probability distribution, and averaging over the values of $O$ 
corresponding to the different configurations. This set of configurations can be generated 
for instance by means of a Monte Carlo procedure, 
or by molecular dynamics \cite{chan-woly81jcp,herm-bern82jcp,parr-rahm84jcp,cepe95rmp}.

By differentiating the partition function with respect to $\beta$, one can 
derive the thermodynamic kinetic energy estimator,
\begin{equation}
\avg{T_{TD}(m)}_m = \avg{\frac{P}{2\beta} -\frac{1}{2P}m\omega_P^2 \sum_{i=0}^{P-1} \left(q_i-q_{i+1}\right)^2}_m.
\label{eq:td-estimator}
\end{equation}
In practice however computing the kinetic energy from Eq.~\eqref{eq:td-estimator} 
is problematic, because the estimator $T_{TD}(m)$
exhibits a large variance that worryingly grows linearly with the number of beads, $P$. 
Hence the statistical error of a simulation of a given length grows
as the number of beads is increased to reach convergence. 
It is however possible to derive a centroid virial estimator \cite{herm-bern82jcp,berne3_1989}, 
\begin{equation}
\avg{T_{CV}}_m = \avg{\frac{1}{2\beta} +\frac{1}{2P} \sum_{i=0}^{P-1} 
           \left(q_i-\bar{q}\right)\frac{\partial V}{\partial q_i}}_m,
\label{eq:cv-estimator}
\end{equation}
where $\bar{q}=\sum_i q_i/P$ is the centroid coordinate. This estimator yields the same
average value as that in Eq.~\eqref{eq:td-estimator} but has a variance that does not grow with $P$. 

\subsection{Isotope substitution by free energy perturbation}\label{sub:fep-estimators}

As shown in Eq. \eqref{eq:iso-a} in order to evaluate the free energy change induced by isotope substitution, one has to 
calculate the change in the quantum kinetic energy upon the change in mass of one of the atoms in the system. 
The most direct approach is to perform a 
series of independent simulations with different values of the isotope mass and compute $\avg{T_{CV}}$ 
for each. Since in most cases one of the 
isotopes has very low natural abundance, this typically involves performing a simulation with a single
particle of mass $\mu$ inside a large supercell containing several tens or hundreds of atoms of the
naturally abundant isotope. This is far from ideal, because one has to perform multiple simulations,
one for each value of $\mu$, in order to evaluate the thermodynamic integration in Eq.~\eqref{eq:iso-a}, and
because statistics are accumulated for just one atom out of the many present in the simulation. 
Hence a highly appealing solution would be to compute $\avg{T_{CV}}_{\mu}$ from a single simulation of \emph{only} the 
most abundant isotope by a free energy perturbation (FEP) method. This approach involves evaluation 
of estimators that simultaneously compute the kinetic energy and correct the phase-space 
distribution to reproduce the statistics of the isotope-substituted system. 

These estimators can be obtained by writing the full expressions for the  
expectation values $\avg{T_{CV}}_{\mu}$ and $\avg{T_{CV}}_{m}$ as phase-space averages. 
By comparing the two expressions, one sees that it is possible to write a ``thermodynamic'' FEP estimator for the kinetic 
energy at mass $\mu$ as the ratio between two phase-space averages computed at mass $m$:
\newcommand{\hTD}[0]{ \ensuremath{h_{\text{TD}}} }
\begin{equation}
\avg{T_{CV}}_{\mu}^{\text{TD}} = 
\frac{
\avg{T_{CV}(\mathbf{q}) \exp\left[- \hTD(\mu/m;\mathbf{q}) \right] }_m
}{
\avg{\exp\left[- \hTD(\mu/m;\mathbf{q}) \right]}_m
} ,
\label{eq:td-fep}
\end{equation}
where we introduce the temperature-scaled difference Hamiltonian
\begin{equation}
 \hTD(\alpha;\mathbf{q})=
\frac{\left(\alpha-1\right)\beta  m\omega_P^2}{2P}  \sum_{i=0}^{P-1} \left(q_i-q_{i+1}\right)^2.
\label{eq:dh-fep}
\end{equation}
These equations provide a recipe to compute the quantum kinetic
energy for an arbitrary mass ratio $\alpha=\mu/m$ out of simulations performed for a single
isotope mass. However, statistical re-weighting comes with its own set of problems which must be 
carefully scrutinized. 
The difference Hamiltonian Eq.~\eqref{eq:dh-fep} appears in the exponent of Eq.~\eqref{eq:td-fep}. 
If it exhibits large fluctuations the weights of different configurations will vary wildly and 
the average performed over many configurations will be dominated by a few outliers,
resulting in very poor sampling efficiency. 
More quantitatively, recent work has shown that whenever one computes re-weighted averages of the 
form $\avg{a e^{-h}}/\avg{e^{-h}}$, the squared error in the mean for an average 
built out of $M$ un-correlated samples will be of the order of $\sigma^2(a) e^{\sigma^2(h)}/M$ \cite{ceri+12prsa}. 
Hence the error grows \emph{exponentially} with the variance of the difference Hamiltonian, $\sigma^2(h)$. 
In addition re-weighted averages are affected by a slowly-decreasing systematic error. 
For re-weighted sampling to be efficient it is therefore necessary for the fluctuations of the 
temperature-scaled difference Hamiltonian to be small and certainly not much larger than one. 

With this in mind Eq.~\eqref{eq:dh-fep} is worrisome in that it is closely reminiscent of 
the thermodynamic kinetic energy estimator in Eq.~\eqref{eq:td-estimator} which is known to exhibit a variance which grows 
with the number of beads, $P$. Due to this similarity we will refer to the estimator in Eq.~\eqref{eq:td-estimator} 
as the TD-FEP estimator. One may therefore wonder whether it is possible to derive an alternative expression which, 
just as the centroid virial kinetic energy estimator in Eq.~\eqref{eq:cv-estimator}, yields the same average 
with a variance which does not grow with $P$. 
To this aim, we use a coordinate-scaling approach in the form of Yamamoto~\cite{yama05jcp}. In our
case, the appropriate scaled coordinates are
\begin{equation}
q_i'(\alpha)=\bar{q}+\frac{1}{\sqrt{\alpha}} \left(q_i-\bar{q}\right).
\label{eq:yama-q}
\end{equation}
This scaling corresponds to a contraction/dilation of the ring 
polymer. Starting from a configuration with a radius of gyration characteristic
of the abundant isotope, it produces one that is more compatible with what would be expected
for an isotope of mass $\mu$. By performing such a change of variables in the 
configurational integral for $\avg{T_{CV}}_\mu$, and then using manipulations analogous 
to those in Ref.~\cite{yama05jcp}, we obtain
\newcommand{\hY}[0]{\ensuremath{h_{\text{SC}}}}
\begin{equation}
\avg{T_{CV}}_{\mu}^{\text{SC}}= 
\frac{
\avg{T_{CV}\left(\mathbf{q}'(\mu/m) \right) \exp\left[- \hY(\mu/m;\mathbf{q}) \right] }_m
}{
\avg{\exp\left[- \hY(\mu/m;\mathbf{q}) \right]}_m
}. 
\label{eq:sc-fep}
\end{equation}
Again, the average is performed in the ensemble that contains only the majority isotope, 
but now the centroid virial estimator is evaluated at the scaled coordinates (Eq.\eqref{eq:yama-q}) and
the temperature-scaled difference Hamiltonian is 
\begin{equation}
 \hY(\alpha;\mathbf{q})=
\frac{\beta}{P}\sum_{i=0}^{P-1} V\left(q'_i(\alpha)\right)-V\left(q_i\right).
\label{eq:yama-dh}
\end{equation}
Unlike the TD-FEP estimator in Eq.~\eqref{eq:dh-fep}, it can be seen that the 
variance of $\hY$ does not grow with $P$. Due to the scaled coordinate transformation
we will refer to this estimator as the SC-FEP estimator. However, unlike the TD-FEP estimator
which can be computed for an arbitrary mass 
ratio in a completely inexpensive way, the SC-FEP estimator requires an 
additional $P$ energy and force evaluations for each mass ratio and tagged particle. 
Such force evaluations can be made inexpensive if the potential consists only in a sum 
of pairwise or otherwise local contributions, because in that case only the interactions 
involving the tagged particle need to be updated. However, in other cases using the 
SC-FEP estimator may entail a significant computational overhead due to the extra potential energy calculations.

Because of the subtle issues connected with re-weighting methods, in order to assess 
the relative merits of these two FEP estimators it is important to perform a careful 
quantitative analysis. 
In  Appendix~\ref{sec:stats} we examine the case of a harmonic potential, where
such analysis can be performed analytically. 
Let us consider here the resulting expressions for the fluctuations of the
difference Hamiltonian, under the further simplifying assumptions that the 
large-$P$ limit is being attained and that we are dealing with a strongly quantized 
normal mode having frequency $\omega_\text{max}$ (i.e. $\beta\hbar\omega_\text{max}\gg 1$). 
For the TD-FEP estimator we find that $\sigma^2(\hTD)$ is proportional to $P$,
as one might have expected:
\begin{equation}
\sigma^2(\hTD)\sim \frac{\left(\alpha-1\right)^2}{2}\left(P-\frac{3}{4} \beta\hbar\omega_\text{max}\right).
\label{eq:var-td}
\end{equation}
In contrast, the difference Hamiltonian for the SC-FEP estimator yields a variance 
which does not grow with $P$, 
\begin{equation}
\sigma^2(\hY) \sim  \frac{\left(\alpha^{-1}-1\right)^2}{2}\left(\frac{\beta\hbar\omega_\text{max}}{4}-1\right).
\label{eq:var-sc}
\end{equation}
As discussed earlier, re-weighting approaches are only reliable when $\sigma^2(h)\lesssim 1$ \cite{ceri+12prsa}. 
This imposes a limit on the range of values of the mass ratio, $\alpha$, for which the TD-FEP 
estimator can be used. The number of beads required to converge a path integral simulation depends on the maximum frequency 
in the system and a reasonable convergence criterion is 
$P = 2\beta\hbar\omega_\text{max}$ \cite{Markland2008}.
Inserting this condition into~\eqref{eq:var-td} yields a constraint on the values of the mass ratio for which the 
TD-FEP estimator is reliable,
\begin{equation}
1-\frac{\sqrt{8/5}}{\sqrt{\beta\hbar\omega_\text{max}}}\le\alpha_{TD}\le 1+\frac{\sqrt{8/5}}{\sqrt{\beta\hbar\omega_\text{max}}}.
\label{eq:const-td}
\end{equation}
Proceeding similiarly for the SC-FEP estimator yields a range of acceptable mass ratios of, 
\begin{equation}
\frac{1}{1+\sqrt{8/(\beta\hbar\omega_\text{max}-4)}}\le\alpha_{SC}\le \frac{1}{1-\sqrt{8/(\beta\hbar\omega_\text{max}-4)}}.
\label{eq:const-sc}
\end{equation}
These analytic estimates allow us to predict how the two approaches might 
compare in a practical case. For example for liquid water at room temperature the 
highest frequency is the OH stretch which extends to frequencies around 3500 ${\rm cm}^{-1}$. 
This gives $\beta\hbar\omega_\text{max}\approx 16$. Inserting this into Eqs.~\eqref{eq:const-td} 
and \eqref{eq:const-sc} yields  acceptable mass ratios of 
\begin{equation}
0.7<\alpha_{TD}<1.3
\label{eq:td-wat-est}
\end{equation}
and
\begin{equation}
0.5<\alpha_{SC}<5.5 
\label{eq:sc-wat-est}
\end{equation}
for the TD-FEP and SC-FEP estimators respectively. Hence the  
TD-FEP approach in Eq.~\eqref{eq:td-fep} would not even allow one to perform a FEP that transforms 
$^1H$ into deuterium ($\alpha=2$), whereas the SC-FEP approach in 
Eq.~\eqref{eq:sc-fep} would be reliable all the way to tritium ($\alpha=3$). 
In the following section we assess the accuracy of these harmonic predictions as to the 
applicability of the FEP estimators and contrast them with the 
direct substitution method for the calculation of free energy changes in liquid water.

\section{Results}

To assess the relative merits of the approaches introduced in the previous section we performed isotope substitution 
simulations in liquid water. Despite its apparent simplicity liquid water provides a challenging test-bed 
since it contains a wide range of frequencies and exhibits a competition between two quantum 
effects, one of which acts to strengthen its hydrogen bonding network and one which weakens it \cite{Chen2003,habe+09jcp}. 
This competition is seen in a variety of other hydrogen bonded systems \cite{li+11pnas} and arises from the 
anharmonicity along the hydrogen bond-covalent bond coordinate \cite{habe+09jcp,mark-bern12pnas}. 
As such water contains many of the features one may encounter in other liquids or hydrogen bonded systems 
and has an important but subtle anharmonic component which must be correctly captured in order to accurately
compute the free energy change upon isotope substitution.

The free energy change we will focus on is the isotopic fractionation of H and D between the 
liquid and the gas phase for water at room temperature defined in Eq.~\eqref{eq:frac-wat} as 
well as the fractionation of $^{16}$O and $^{18}$O. The results are quoted as fractionation 
factors as defined in Eq.~\eqref{eq:frac-fac} which correspond to  the ratio of isotope in the liquid to that in the vapor.
The fractionation of isotopes is a frequently used technique to understand processes as diverse as the water 
cycle in the earth's climate, 
biological function and the origins of the interstellar medium \cite{Worden2007,Berner03032000,Paniello:2012fk}. 

As can be seen from Eqs.~\eqref{eq:iso-a} and \eqref{eq:frac-wat} the liquid-vapor fractionation 
ratio probes the change in the quantum kinetic energy upon moving a water molecule from the 
liquid phase to the gas phase. An increase in kinetic energy occurs when a quantum particle is confined by the forces 
exerted on the particle by the surrounding molecules. Computing fractionation ratios is particularly challenging, 
since their precise value 
depends on subtle cancellations of different quantum effects. As such, they could 
serve as a sensitive test of how well different potentials reproduce this delicate 
balance.

We performed our simulations using the flexible simple point charge q-TIP4P/F water 
model \cite{habe+09jcp}. This model was chosen as it has been shown to accurately predict 
the H/D fractionation between liquid water and its vapor \cite{mark-bern12pnas} 
as well as a variety of other properties of water such as its density maximum and melting 
point \cite{habe+09jcp} when used in PIMD simulations.  

We will compare three different 
approaches for computing the free energy change upon isotope substituion: 1.) the previously 
introduced direct subtitution approach; 2.) the TD-FEP estimator in Eq.~\eqref{eq:td-fep} 
and 3.) the SC-FEP estimator in Eq.~\eqref{eq:sc-fep}.

\subsection{Generating path integral configurations}
\label{sec:gen-pi}

Evaluating the expectation values of the kinetic energy in Eq.~\eqref{eq:cv-estimator} or 
the re-weighted kinetic energies in Eqs.~\eqref{eq:td-fep} and \eqref{eq:sc-fep} 
requires the generation of path integral trajectories. We will compare two approaches: 
an efficiently thermostatted implementation of PIMD \cite{ceri+10jcp} and the recently 
introduced PIGLET method \cite{ceri+11jcp}.

The difference in these approaches is in their convergence with respect to the 
number of beads employed. In conventional PIMD \cite{parr-rahm84jcp} in order to converge a simulation 
for a system whose fastest vibrational mode has a 
frequency $\omega_\text{max}$, the number of beads, $P$, must be two or more 
times the parameter $\beta\hbar\omega_\text{max}$. For water at room 
temperature typically $P=32$ yields results with an error of a few percent. 
In most simulations  the computational effort is dominated by the calculation of the forces, 
and so the total cost of a conventional PIMD simulation scales 
linearly with beads -- hence making the PIMD simulation $\sim32$ times more expensive than 
a classical one.  The PIMD simulations were performed using the PILE-G thermostatting scheme \cite{ceri+10jcp} 
that has previously been demonstrated to be an efficient approach to simulate liquids. 
Since the global thermostat minimally hinders diffusion we used a 
short thermostat correlation time of $\tau_\text{thermo}=25$~fs.

\begin{figure}
\caption{\label{fig:p-converge} a) Expectation values for the centroid-virial kinetic energy estimator
for hydrogen in liquid water as a function of the number of beads. Results are shown for both
PIMD and PIGLET simulations, and for the tagged particle having mass set to that of hydrogen ($m_H$) 
or deuterium ($m_D$). 
b) Convergence of the isotope fractionation ratio $\alpha_{l-v}$ between
the liquid and the gas phase of water for the H/D substitution. Results are reported
as a function of the number of beads, using conventional path integral MD (blue circles)
and PIGLET (red lozenges). Lines are just guides for the eye, and 
statistical error bars are smaller than the size of the points.}
\includegraphics[width=1.0\columnwidth]{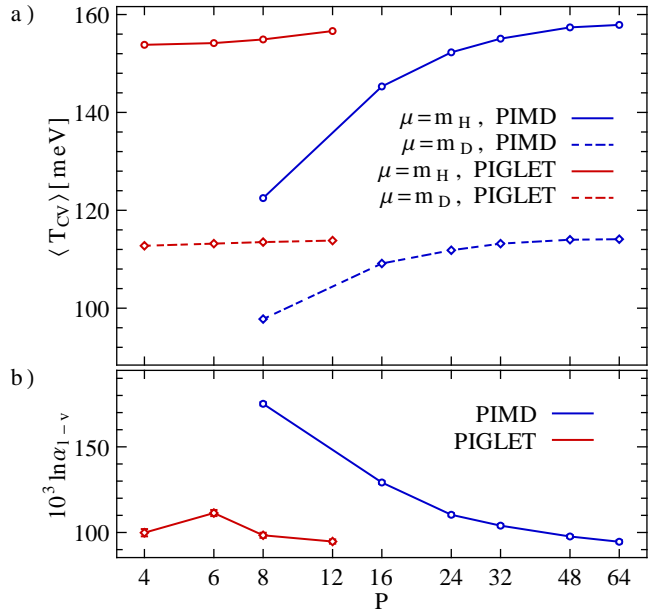}
\end{figure}

Recently introduced colored-noise Langevin dynamics approaches can reduce the cost of 
path integral simulations by enforcing certain pre-determined bead-bead correlations 
to be exact in the harmonic limit~\cite{ceri+11jcp}. This dramatically reduces the number of beads 
needed to converge many properties, even in anharmonic systems. 
Of these colored-noise approaches the PIGLET method \cite{ceri-mano12prl} is particularly relevant 
to the present work since it was designed to ensure fast convergence of the quantum kinetic energy, 
$\avg{T_{CV}}_\mu$ which is all that is required to compute the free energy change by explicit isotope 
substitution. The fast convergence of the kinetic energy using PIGLET is demonstrated in 
Figure~\ref{fig:p-converge}a. Using PIMD with $P=32$ yields an estimate of the kinetic energy 
with an error of about 2\%. PIGLET yields results with similar accuracy to this using 
only 6 beads. Figure~\ref{fig:p-converge}b shows the fractionation ratio evaluated from Eq.~\eqref{eq:frac-fac}.  
This property is the small difference of two large quantities and 
is therefore very sensitive to errors in the description of anharmonicities, 
which is apparent in the very poor performance of the quasi-harmonic approximation (see Sec.~\ref{sub:harm}).
Nevertheless, PIGLET performs remarkably well: from 4 to 12 beads, the values of 
$\ln \alpha_{l-v}$ differ by less than 10\% from the asymptotic value, an error which is much lower
than the one stemming from the use of an empirical potential \cite{mark-bern12pnas}.

However, even in the harmonic limit,
the PIGLET approach does not enforce all the bead-bead correlations 
needed for the expectation values of the TD-FEP and SC-FEP estimators in 
Eq.~\eqref{eq:td-fep} and Eq.~\eqref{eq:sc-fep}. 
Therefore, there is no guarantee that PIGLET will converge much faster than 
conventional PIMD when used together with one of the FEP estimators.  
On the other hand, the earlier  PI+GLE colored noise approach \cite{ceri+11jcp} was 
shown to enhance convergence of the kinetic energy even though it was not 
developed to enforce all the required correlations. This suggests that PIGLET, even in its 
current form, might be applied to calculate the FEP estimators as we investigate in Secs.~\ref{sec:td-fep} and \ref{sec:sc-fep}.

\subsection{Direct computation of  $\avg{T_{CV}}_{\mu}$ }\label{sub:direct}

The direct technique to compute isotope effects involves evaluating the kinetic 
energy $\avg{T_{CV}}_{\mu}$ in multiple simulations where the mass of one of the 
atoms is changed between those of the two isotopes in order to perform the 
thermodynamic integration in Eq.~\eqref{eq:iso-a}.  As demonstrated 
in Sec.~\ref{sub:th-int}, with a physically motivated change of integration variable 
it is possible to accurately evaluate the integral in Eq.~\eqref{eq:ti-transformed} by 
computing just the extremal points. Despite the need to replicate the system 
this can be much cheaper than simply using a quasi-harmonic approximation 
from a classical trajectory.

One reason for this is the efficiency in de-correlating the kinetic energy in 
a path integral simulation. The statistical uncertainty of a property computed
out of a simulation of length $t_\text{sim}$ decreases in proportion to 
$\sqrt{\tau/t_\text{sim}}$ where $\tau$ is the correlation time 
of that property. Hence, for a given length of the simulation, observables with 
short correlation times will exhibit smaller statistical error.
In this regard the PILE-G thermostat we used for PIMD calculations 
and PIGLET are both very efficient. Our simulations show the autocorrelation 
time of the centroid-virial kinetic energy estimator to depend weakly on 
$P$ and $\mu$, varying between $1.5$ and $2$~fs for all the simulations 
we performed. We used a time-step of $0.25$~fs and hence $6-8$ force 
evaluations are enough to obtain a configuration where the kinetic energy estimator is uncorrelated. 
In addition, because of the small number of beads required 
for convergence, PIGLET simulations could have been performed with a $0.5$~fs time-step,
which enhances the efficiency even further. 
This explains why despite the additional cost of path integral approaches in terms of 
force evaluations they are still a cheaper approach than using the quasi-harmonic 
approach in Sec. \ref{sub:harm}.
In our case we used an inexpensive empirical potential, and therefore we ran simulations 
that were as long as 500~ps, resulting in minuscule statistical errors. 
If one was prepared to accept a $\pm 5$ uncertainty 
on the H/D fractionation ratio, liquid water simulations of the order of 10~ps 
starting from an equilibrated configuration would suffice, which makes 
performing {\it ab initio} fractionation calculations for liquids highly plausible.

\begin{figure}
\caption{\label{fig:size-converge} Average kinetic energy (blue circles) computed from simulations
in which $n_D$ hydrogen atoms have been substituted with deuterium,
in a simulation box containing 64 water molecules. At most one atom per molecule was substituted.}
\includegraphics[width=1.0\columnwidth]{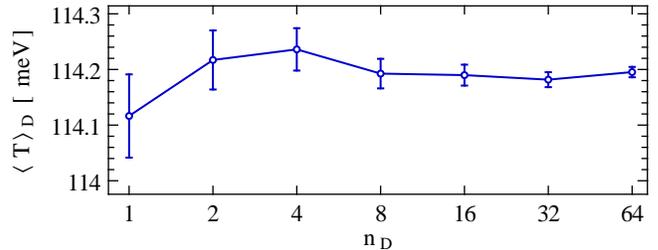}
\end{figure}

Due to their naturally low abundance, fractionation ratios are typically required in the 
low-dilution limit with only trace concentrations of the minority isotopes. In order to 
simulate this limit exactly the mass has to be changed for a single atom in a system that 
contains hundreds of instances of the naturally-abundant isotope, and the kinetic energy 
can be computed for that particle only. However, in a homogeneous system such as water 
it is interesting to explore how sensitive the kinetic energy, and hence the computed 
fractionation ratio, are to multiple isotope substitutions in the system. The advantage 
of making these multiple isotope substitutions is that one can then gather statistics from 
several atoms at the same time. Momentum-dependent observables tend to be very local in nature, 
and one can hope that the error introduced by having an unnaturally high concentration of isotope substituted
species will be small. For instance, when performing open-path PIMD in liquid water, one can
simultaneously open the path for one hydrogen atom per water molecule without introducing significant
systematic errors \cite{morr+07jcp,morr-car08prl}. Figure~\ref{fig:size-converge} shows how $\avg{T_{CV}}_{m_D}$
changes when multiple atoms in a simulation of light water are transmuted into deuterium, with the precaution of never 
substituting two atoms on the same molecule. Even when half the H's in the system are converted, 
the kinetic energy of the D's changes by less than 0.3\%, a difference that is not statistically 
significant even with these relatively long calculations. 
When performing more demanding \emph{ab initio} calculations, the statistical error from the short runs will almost certainly
be dominant, making it particularly attractive to gather horizontal statistics by multiple 
substitutions. One can also observe that when multiple substitutions are performed the statistical uncertainty
of $\avg{T_{CV}}_{\mu=m_D}$ decreases with the square root of the number of substituted atoms. 
This implies that the instantaneous values of $T_{CV}$ for atoms in 
different molecules are almost perfectly un-correlated -- another indication of the local nature of the
quantum kinetic energy.
However, we stress that one should be careful when performing multiple substitutions, in particular
for inhomogeneous systems, and that whenever possible one should benchmark this approximation 
against the highly-diluted case. 

Even exploiting our efficient transformation of the integral which requires 
only simulations at the two extrema the need to perform multiple simulations to obtain 
free energy changes using direct substitution is undesirable. We now consider our FEP 
approaches which allow the free energy change to be obtained from a single trajectory 
of the most abundant isotope.

\subsection{``Thermodynamic'' free energy perturbation}
\label{sec:td-fep}

\begin{figure}
\caption{\label{fig:td-fep-cumul} Cumulative average of $\avg{T_{CV}}_{\mu}^{TD}$, as computed for 
hydrogen isotopes in a single PIMD trajectory of liquid water, with $P=32$.
Lines, from top to bottom, correspond to different mass ratios ranging from $\mu/m_H=1$ to $\mu/m_H=2$.}
\includegraphics[width=1.0\columnwidth]{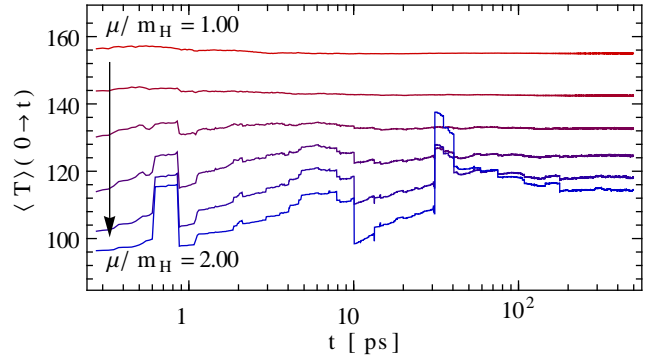}
\end{figure}

The TD-FEP estimator in Eq.~\eqref{eq:td-fep} provides an appealing
approach to the computation of the kinetic energy of an isotope-substituted system by just
post-processing the data from a simulation that contains only the naturally-abundant
isotope. For a homogenous system such as water this ``virtual substitution'' can be performed 
one atom at a time for every atom in the system allowing full horizontal statistics to be 
collected. However, as discussed in Sec.~\ref{sec:theory} the re-weighting procedure which underlies 
free-energy perturbation methods is potentially dangerous, as it can introduce systematic errors and a 
difficult-to-detect statistical inefficiency \cite{ceri+12prsa}. 
The presence of unusual statistical behavior, dominated by outliers, is demonstrated in Fig.~\ref{fig:td-fep-cumul},
which shows the cumulative average of the kinetic energy predicted from Eq.~\eqref{eq:td-fep} for 
different mass ratios as the simulation progresses. For large values of $\mu/m_H$ the cumulative average does not converge 
with regularity to a mean value, but exhibits a characteristic sequence of plateaus, 
interrupted by abrupt jumps when a new outlier is encountered. 

\begin{figure}
\caption{\label{fig:td-fep-deltah} Contour plot of the variance of the difference Hamiltonian
for the thermodynamic free energy perturbation $h_{TD}$, as a function of the number of beads
$P$ and the mass ratio $\alpha=\mu/m_H$. For $\sigma^2(h_{TD})\gtrsim 1$ computing $\avg{T_{CV}}_{\mu}^{TD}$
becomes impractical.}
\includegraphics[width=1.0\columnwidth]{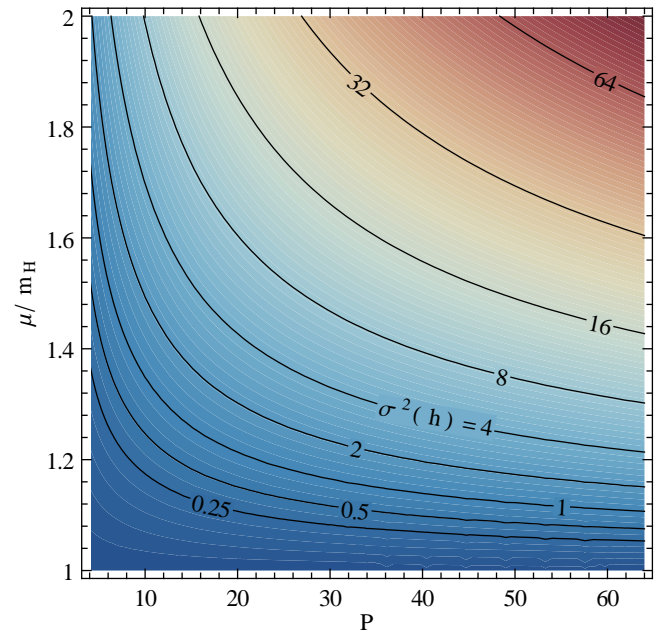}
\end{figure}

A more quantitative approach to monitor the statistical quality of the TD-FEP estimates of $\avg{T_{CV}}_{\mu}^{TD}$
involves computing the variance of the temperature scaled difference Hamiltonian in Eq.~\eqref{eq:dh-fep}.
As discussed in Sec.~\ref{sec:theory} a variance, $\sigma^{2}(h)$, smaller than one is required for 
re-weighting to be of practical use. Figure~\ref{fig:td-fep-deltah} shows the variance obtained from 
our water simulations for different combinations of the number of beads, $P$ and mass ratio, $\alpha$. 
As expected from our analysis in Appendix~\ref{sec:stats} the variance of the TD-FEP 
grows rapidly with both $P$ and $\alpha$ so in practice this estimator becomes useless for mass 
ratios greater than about $1.2$. Importantly the simulated behavior of the variance is in excellent 
agreement with the analytic prediction derived for the harmonic limit in Eq. \eqref{eq:td-wat-est} 
which predicts a maximum mass ratio of $1.3$ for water simply from the knowledge of the highest 
frequency present. Hence, the analytic estimates provide an excellent {\it a priori} way of deciding whether one should employ TD-FEP for a given system.

\begin{figure}
\caption{\label{fig:iso-oxygen} Convergence of the isotope fractionation ratio $\alpha_{l-v}$ between
the liquid and the gas phase of water at ambient conditions 
for the $^{16}$O$/$$^{18}$O substitution. Upper panel depicts the convergence
of the $^{18}$O kinetic energy in the two phases, as computed from thermodynamic free energy perturbation,
while the lower panel reports the fractionation ratio. Lines are just guides for the eye.}
\includegraphics[width=1.0\columnwidth]{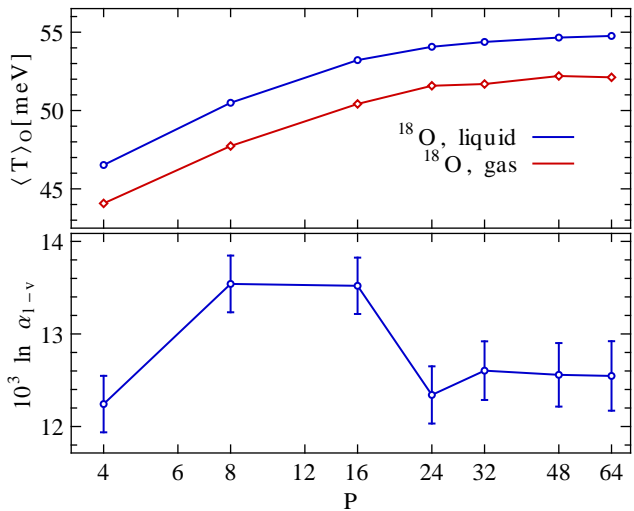}
\end{figure}

A maximum mass ratio of $1.2$ using conventional PIMD simulations implies that TD-FEP is not recommended for studying isotope substitution 
for hydrogen in room temperature water but this mass ratio is within the range needed to perform 
$^{16}$O$/$$^{18}$O substitution, which involves a mass ratio of only $\alpha=18/16$. 
Figure~\ref{fig:iso-oxygen} shows the isotope fractionation ratio for $^{16}$O and $^{18}$O between 
the liquid and the gas phases of water at ambient conditions. The isotope
substitution free energies were computed by TD-FEP from a single simulation 
of pure $^{16}$O water. Despite being a heavy atom the number of beads required to converge 
the kinetic energy in the liquid and gas phase is $\sim32$ since it is bonded to a light atom. The fractionation ratio converges
with the number of beads more rapidly than the kinetic energy estimator does, because of a 
cancellation of errors between the liquid and the gas phases. 
We find $10^3\ln \alpha_{l-v}=12.5\pm0.5$, for the q-TIP4P/F model which is
in reasonable agreement with the experimental value of 
$9.2$ \cite{Horita1994}. This is in error by an amount similar to a fall of 30 K in 
the temperature which is roughly the same as the error in the H/D liquid vapor fractionation when 
compared to experiment \cite{mark-bern12pnas} suggesting this model mildly over-predicts fractionation in water.

\begin{figure}
\caption{\label{fig:piglet-fep-td} 
The figure demonstrates the performance of the TD-FEP estimator used together with PIGLET.
Panel a) represents $\avg{T_{CV}}_{m_D}$ as a function of the number of beads. PIMD results are shown for comparison,
and the black line indicates the reference value (direct substitution, PIMD, $P=64$). 
In Panel b) the isotope fractionation ratio between liquid and vapor is shown, for water at room temperature, 
as estimated from TD-FEP and PIGLET. The black line indicates the reference value (direct substitution, PIMD, $P=64$).
}
\includegraphics[width=1.0\columnwidth]{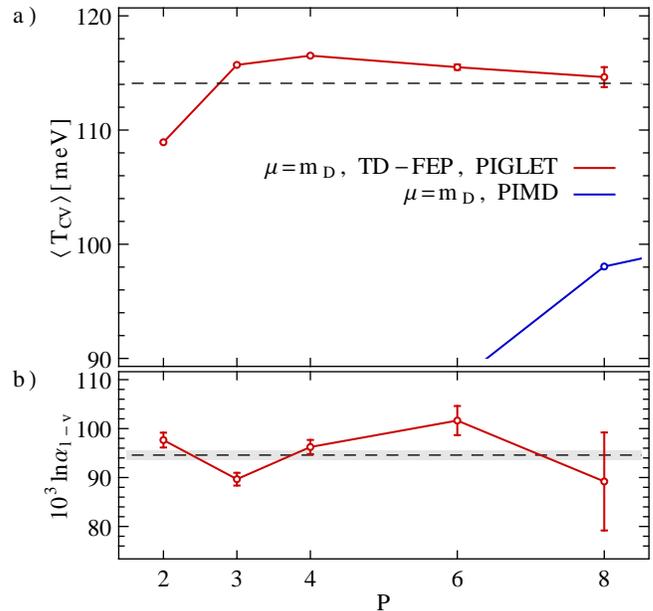}
\end{figure}

As we discussed in Sec. \ref{sec:gen-pi} PIGLET does not enforce the bead correlations necessary to ensure that 
it will accelerate the convergence of the TD-FEP estimator even in the harmonic limit. However, 
there are two reasons why such a combination could be highly beneficial and hence is worth investigating. 
The first is simply the reduction in the number of beads, which implies that fewer force evaluations 
are required to evolve the system resulting in a speed-up of $\sim6$ times in water. 
The second is that the variance of the difference Hamiltonian that exponentially determines 
the rise in the error of TD-FEP depends on the number of beads (see Fig. \ref{fig:td-fep-deltah}). 
Hence any reduction in the number of beads by using PIGLET would offer exponential benefits in the efficiency of the TD-FEP approach.

Figure~\ref{fig:piglet-fep-td}a shows the value of the kinetic energy of $\avg{T_{CV}}_{m_D}$
estimated from a single simulation of pure liquid H$_2$O, using TD-FEP with PIGLET. 
Without PIGLET, at least 32 beads would be needed to converge the kinetic energy, which as discussed 
above would restrict the maximum acceptable mass ratios to about $1.2$. 
On the other hand, as shown in Fig.~\ref{fig:piglet-fep-td}b, PIGLET reaches convergence with about 6 beads 
for both the kinetic energy and the fractionation ratio, which is just on the boundary of the region 
for which performing TD-FEP up to $\alpha=2$ would become impractical.
The fact that PIGLET was not parameterized to enforce the particular correlation needed for TD-FEP does not 
appear to limit its utility in converging it and indeed remarkably it even seems to converge as rapidly 
as kinetic energy, which is rigorously enforced in the harmonic limit.
Although by reducing the required number of beads PIGLET delays the failure of TD-FEP 
the statistical error blows up rendering the estimator unusable beyond 6-8 beads (see Figure~\ref{fig:td-fep-deltah}). 
The extra window of mass ratio PIGLET opens is however incredibly useful since it allows H/D isotope 
substitution to be computed at room temperature and for ubiquitous aqueous and hydrogen-bonded systems,
performing just a single trajectory and using as few as 6 beads. 

\subsection{Scaled coordinates free energy perturbation}
\label{sec:sc-fep}

In the previous section TD-FEP was shown to be limited due to possessing a statistical error 
that grows with the number of beads. Based on our harmonic analysis the SC-FEP approach 
should allow one to obtain a much wider range of mass ratios $\alpha$ from a single trajectory. 
In addition the analytic predictions suggest that, unlike TD-FEP, the variance of the difference Hamiltonian for SC-FEP will show no dependence on the number of beads allowing the path integral simulation to be performed to high precision without the statistical inefficiency rising to unacceptable levels. 

\begin{figure}
\caption{\label{fig:sc-fep-deltah} Contour plot of the variance of the difference Hamiltonian
for the coordinate-scaling free energy perturbation $\hY$, as a function of the number of beads
$P$ and the mass ratio $\alpha=\mu/m_H$. Throughout the range we considered, $\sigma^2(h_{SC}) < 1$. }
\includegraphics[width=1.0\columnwidth]{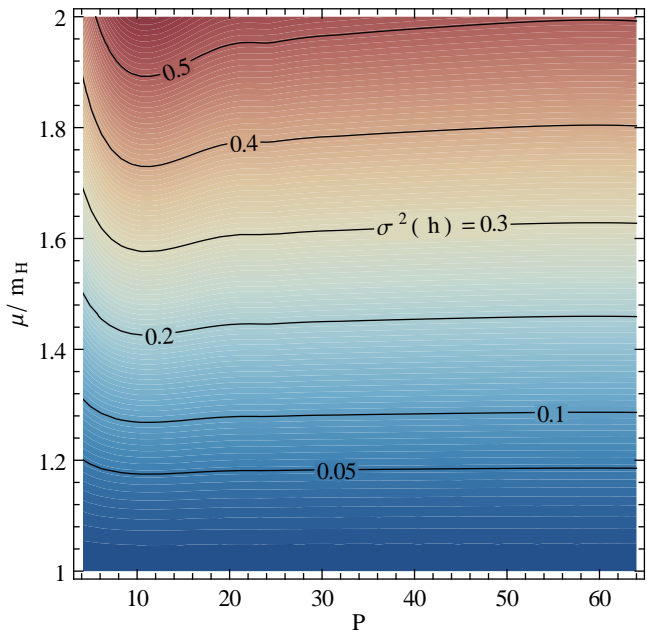}
\end{figure}

To test these predictions Fig.~\ref{fig:sc-fep-deltah} shows the variance of the difference Hamiltonian $\hY$ for SC-FEP as a function of 
$P$ and $\mu/m_H$, over the same range of masses and numbers of beads we explored
in Figure~\ref{fig:td-fep-deltah} for TD-FEP. In striking contrast with the case of the TD-FEP,
$\sigma^2(h_{SC})$ is nearly independent of $P$, and even up to a mass ratio of $\alpha=2$ is well below the threshold value of one. The SC-FEP estimator is therefore suitable for computing H/D substitution using conventional PIMD using either $P=32$ or $64$ beads depending on the accuracy required. Based on the 
scaling of the variance with temperature predicted by Eq.~\eqref{eq:const-sc} one would 
expect this to be the case even at temperatures considerably lower than 300~K. 

\begin{figure}
\caption{\label{fig:piglet-fep-sc} 
The figure demonstrates the poor convergence of the SC-FEP estimator when used together with PIGLET.
a) $\avg{T_{CV}}_{m_D}$ as a function of the number of beads. PIMD results are shown for comparison,
and the black line indicates the reference value (direct substitution, PIMD, $P=64$). 
b) Isotope fractionation ratio between liquid and vapor for water at room temperature, estimated from
SC-FEP and PIGLET. The black line indicates the reference value (direct substitution, PIMD, $P=64$).
}
\includegraphics[width=1.0\columnwidth]{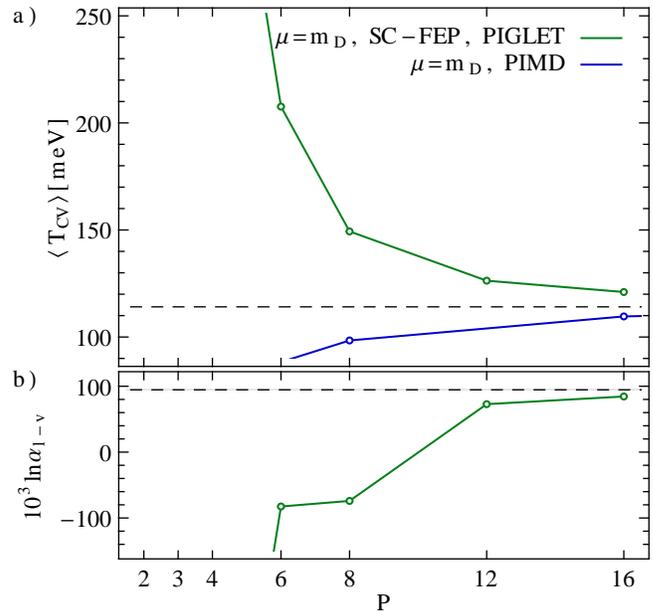}
\end{figure}

Given SC-FEP's much greater range of applicability than TD-FEP when used in conjunction with conventional PIMD one might wonder if SC-FEP could be reduced in cost by using PIGLET which was shown to be highly effective in accelerating convergence of TD-FEP. 
Figure~\ref{fig:piglet-fep-sc} demonstrates that in this case the bead-bead correlations needed to converge the SC-FEP estimator
are clearly not included in the current PIGLET parameterization leading to very poor convergence. When using PIGLET with 6 beads, which is essentially converged for TD-FEP or direct substitution,  the SC-FEP estimator gives a kinetic energy
estimate for deuterium with an error of 100\% and a fractionation ratio which does not even possess the correct sign. 
It is reassuring to note that, by further increasing the number of beads, convergence will 
be eventually reached.  This underlines the robustness of the PIGLET approach: when a 
sufficiently large number of beads are used, the non-equilibrium nature of the colored noise 
naturally vanishes and these methods reduce to conventional PIMD. 
It is also worth mentioning that in principle one could exploit the philosophy of 
the PIGLET method to enforce in the harmonic limit the bead-bead correlations that
are necessary to rigorously converge both TD-FEP and SC-FEP. 
Such a re-parameterization of the colored noise is a non trivial task and hence is left for future work.

\subsection{Recommendations}

From the discussion above it is clear that each approach for obtaining the  
free energy change upon isotope exchange has benefits and deficits. As such there can be no single recommendation 
but rather, based on our simulations and analytic results, one can suggest the most appropriate 
approach to efficiently tackle typical simulation scenarios. 

The direct substitution approach provides a baseline since it has been extensively used in previous 
studies \cite{Vanicek2007,Paesani2009,Perez2011,mark-bern12pnas}. In the present paper we demonstrate a number of stratgies 
that greatly improve its efficiency. For a start, by performing a physically motivated smoothing of the integral 
(Section~\ref{sub:th-int}) one can usually perform the thermodynamic integration with only two simulations for the 
extremal values of the isotope mass. In addition the PIGLET approach can be used to reduce the number of force evaluations 
required for each simulation. Finally, our results suggest that in a homogeneous system 
performing multiple substitutions is a reliable way to obtain horizontal statistics -- although this 
should of course be treated with caution and tested carefully before it is applied to other problems.  
With these improvements, direct substitution becomes a very competitive choice even in comparison to approximate methods.  

The free energy perturbation approaches we introduce, however, are even more appealing, in that
they allow one to obtain the isotope substitution free energy from a single simulation containing only the
 majority isotope, and allow one to gather horizontal statistics in an exact way. However, one must be careful 
of the statistical efficiency of re-weighting whenever large fluctuations occur in the difference Hamiltonian.
 In the preceding sections we have presented both analytical and simulation evidence to determine when FEP 
approaches are effective. This allows us to provide some simple practical guidelines:
\begin{itemize}
\item For heavy isotope substitution at temperatures above 100K, one should perform a PIGLET simulation 
containing only the most abundant isotope and use the TD-FEP estimator to extract the free energy change.
\item For H/D substitution in ``biological'' conditions (room temperature, aqueous systems) and at higher 
temperatures, one should also use TD-FEP and PIGLET. However, this more challenging regime sits on the edge 
of the TD-FEP estimator's applicability when combined with PIGLET. Hence TD-FEP should not be used with 
PIMD in this regime since the additional beads required for convergence cause the statistical efficiency 
to become unacceptable.
\item For isotope substitution below room temperature for H/D or at room temperature when making a 
substitution with a larger mass ratios than H/D (e.g Muonium/H or H/Tritium) a direct substitution 
calculation is the best option. The efficiency of direct substitution can be greatly enhanced by using 
our transformed integration variable, PIGLET and, where appropriate, using multiple substitutions. 
\end{itemize}

While the three cases above cover most practical applications, there are circumstances in
which the SC-FEP estimator becomes competitive. Due to the requirement to evaluate the potential 
on a scaled ring polymer its main advantage lies in cases where the potential can be evaluated 
efficiently upon a single particle change such as for empirical potentials. As we demonstrated the 
convergence of the SC-FEP is not improved by PIGLET and so PIMD must be used to simulate the trajectory. 
However for empirical potentials PIMD trajectories can be made extremely cheap by using ring-polymer 
contraction techniques\cite{Markland2008,mark-mano08cpl,Fanourgakis2009}. SC-FEP is also more attractive 
in problems where only a single atom in the system is substituted such as for a particular proton in 
an enzyme active site. 

Since our analytic predictions of the fluctuations in the difference Hamiltonians for the FEP 
estimators were seen to be reliable when compared to liquid water simulations we can use them to 
derive simple expressions that predict the efficiency of TD-FEP and SC-FEP relative to direct simulation. 
These efficiency measures, $E_{TD}$ and $E_{SC}$ gauge the cost of obtaining the free 
energy change from the TD-FEP and SC-FEP estimators respectively compared to obtaining it from direct 
substitution. If $E_{TD}$ or $E_{SC}$ come out much greater than one then the FEP approach is preferable and 
the FEP variant with the highest value should be used. If both values come out much less than one 
then direct substitution should be employed. Values around one suggest that either FEP or direct 
substitution are of comparable efficiency for the problem. For an isotope substitution with a mass 
ratio $\alpha=m'/m$ between the most and least abundant isotope at inverse temperature $\beta$ in a 
system with a maximum vibrational frequency $\omega_\text{max}$ one can show that for the TD-FEP 
estimator the efficiency is
\begin{equation}
E_{TD} = n_\text{int} \sqrt{\frac{N_{X}}{M_{X}} } 
\exp\left[-\frac{\left(\alpha-1\right)^2}{4}\left(P-\frac{3}{4} \beta\hbar\omega_\text{max}\right)\right]. 
\label{eq:eff-td} 
\end{equation} 
Here $n_\text{int}$ is the number of integration points one needs to converge the thermodynamic integration
(2 or 3 when using Eq.~\eqref{eq:ti-transformed}), $N_{X}$ is the number of equivalent atoms for which 
the substitution can be performed, and $M_{X}$ the number of multiple substitutions one can perform 
in a direct calculation ($M_X=1$ is the only completely safe bet). $P$ is the number of beads required for convergence, which 
is of the order of $2 \beta\hbar\omega_\text{max}$ with conventional PIMD and typically about 
$\beta\hbar\omega_\text{max}/3$ with PIGLET. From this equation one can immediately see that for 
systems which require large numbers of integration points, $n_{int}$ or posses a large number of 
exchangeable atoms then TD-FEP holds an advantage. However, at low temperature (high $\beta$) 
or a large mass ratio or high frequencies $\omega_\text{max}$ the efficiency can quickly shift in favour 
of direct substitution. One can also immediately see why PIGLET is so useful when used with TD-FEP 
since the relative efficiency is exponentially sensitive to the number of beads used so, by reducing 
the number of beads required by a factor of $\sim6$, PIGLET yields a $e^{6}\approx 400$ times 
increase in $E_{TD}$. 

The efficiency of the SC-FEP estimator is,
\begin{align}
 E_{SC} =  n_\text{int}  &\sqrt{\frac{\tau N_{X} P_\text{GLE} }{\Delta t(1+K N_{X} ) M_{X} P} } \times \nonumber \\
 &  \exp\left[-\frac{\left(\alpha^{-1}-1\right)^2}{4}\left(\frac{\beta\hbar\omega_\text{max}}{4}-1\right)\right].
    \label{eq:eff-sc}
\end{align}
Besides the quantities present in Eq.~\eqref{eq:eff-td}, this expression also contains $\tau$,
the correlation time of $T_{CV}$, $\Delta t$, the time step of the simulation when using PIGLET, 
and $K$, the cost of computing the scaled-coordinates potential relative to the cost 
of a simulation step (e.g. $K=1$ for {\it ab initio}, and virtually $K=0$ if the potential can
be evaluated locally). Since SC-FEP cannot be used together with PIGLET, the expression
also contains both the number of beads required to converge PIGLET, $P_\text{GLE}$ and 
the number of beads required to converge conventional PIMD, $P$. Here it can be seen that for 
systems with long correlation times $\tau$, cheap scaled-coordinate potential evaluations $K$ or 
low numbers of exchangeable sites $N_{X}$, SC-FEP dominates.

\section{Conclusions}

In this paper we have introduced a series of improvements which enhance the ease and efficiency of calculating 
isotope exchange free energies. This include two new estimators, based on free energy perturbation 
techniques as well as a number of improvements to the existing direct substitution approach. We then critically discussed 
the advantages and shortcomings of these approaches predicating our reasoning on both analytic arguments
and on extensive simulations of room-temperature water. 

The balance of pros and cons is fundamentally related to the statistical efficiency of sampling
of the different techniques. This, in turns, depends on the ``quantumness'' of the system, as 
expressed by the ratio between the highest quantum of vibrational energy and the thermal energy,  
$\beta\hbar\omega_\text{max}$  and on the 
ratio of the masses of the minority isotope and of the naturally abundant one.

When this mass ratio is close to one, as is the case whenever one is dealing with relatively heavy elements
such as oxygen, a ``thermodynamic'' FEP estimator allows one to compute inexpensively the effect of isotope 
substitution based on a single simulation which only contains the majority isotope. This both simplifies and makes
less computationally demanding the evaluation of quantities of great interest for geochemistry and
climatology. 

When instead the mass ratio is large, such as when one needs to substitute deuterium for hydrogen,
or when the temperature is considerably below room temperature, the scenario is less clear-cut and 
one has to weight carefully  a number of factors. 
The statistical efficiency of the ``thermodynamic'' estimator
degrades dramatically with the number of beads, so it becomes mandatory to use PIGLET to ensure that 
convergence is achieved with a reduced number of beads. 
With this caveat, H/D substitution is still manageable with 
TD-FEP in most systems of interest at room temperature. 
At cryogenic temperatures an alternative estimator can be used,
which is based on a scaling of coordinates relative to the centroid of the path and 
exhibit more robust statistical properties. However it requires multiple evaluations 
of the forces, and is therefore more computationally demanding than TD-FEP. 
For simulations using empirical potentials it is possible to reduce this computational
cost by exploiting the locality of the coordinate scaling procedure. For an \emph{ab initio}
simulation, however, there might be corner cases in which a carefully-performed
direct substitution simulation is more advantageous that either TD-FEP and SC-FEP. 

We provide analytical expressions that allow one to estimate \emph{a priori} the 
different factors that determine the sampling efficiency, so that one can make the optimal 
choice for a given system without having to perform time-consuming benchmarks. 
This analysis and the estimators we introduce will greatly simplify the evaluation of observables 
connected with isotope substitution, that can be measured experimentally with great precision
and provide precious insight into a multitude of natural phenomena. Obtaining
the corresponding theoretical predictions will be useful both to elucidate and predic the results of experiments, and to validate and benchmark empirical and {\it ab initio} predictions of condensed-phase systems. 

\section{Acknowledgments}
We would like to thank David Manolopoulos for insightful discussion and
many precious suggestions and Timothy Berkelbach for a thorough reading of the manuscript.
M.C. acknowledges funds from the EU Marie Curie IEF No. PIEFGA-2010-272402. 
T.E.M acknowledges funding from a Terman fellowship and Stanford start-up funds.

\appendix

\section{Reweighting statistics for FEP estimators}
\label{sec:stats}

When developing new estimators in a path integral context, analysing their 
properties in the harmonic limit often allows one to obtain analytic results
and useful insight into their limitations.
This is particularly important when performing re-weighted sampling, 
because of the risk of dramatically degraded sampling statistics. 
In what follows we will consider a one-dimensional harmonic oscillator
of frequency $\omega$ and mass $m$, simulated at temperature $k_B/\beta$ and using
$P$ beads.
Consider for instance the expectation value of the difference Hamiltonian for the 
TD-FEP estimator~\eqref{eq:dh-fep}.  
One can first transform into  the normal-modes coordinates of the path 
$\tilde{q}_k$ \cite{bern-thir86arpc}, and get 
\begin{equation}
\avg{\hTD}_m= \frac{(\alpha-1) m\beta}{2P}\sum_k \omega_k^2\avg{\tilde{q}_k^2}_m
\end{equation}
One can then compute analytically the average fluctuations of the normal mode 
coordinates, and take the limit of an infinite number of beads analytically, 
which yields
\begin{equation}
\begin{split}
\avg{\hTD}_m&=\frac{\alpha-1}{2} \left(P-\sum_k \frac{1}{1+\omega_k^2/\omega^2} \right) 
\underset{P\rightarrow\infty}{=} \\
&=\frac{\alpha-1}{2} \left(P-\frac{x}{2}\coth\frac{x}{2}\right).
\end{split}
\end{equation}
Here we introduced the frequency of the $k$-th normal mode of the free ring polymer,
 $\omega_k=2\omega_P \sin k\pi/P$, and the shorthand $x=\beta\hbar\omega$.
Proceeding in a similar way, one can obtain the fluctuations of $\hTD$,
\begin{equation}
\begin{split}
\sigma^2(\hTD)&=\left[\frac{(\alpha-1) m\beta}{2P}\right]^2 2 
\sum_k \left[\omega_k^2\avg{\tilde{q}_k^2}_m\right]^2=\\
&=\frac{(\alpha-1)^2}{2} \left(P+\sum_k 
\frac{1-2\left(1+\omega_k^2/\omega^2\right)}{\left(1+\omega_k^2/\omega^2\right)^2}\right)\underset{P\rightarrow\infty}{=} \\
&=\frac{(\alpha-1)^2}{2} \left[P-\frac{3}{4}x\coth \frac{x}{2} +\frac{1}{2}\left(\frac{x/2}{\sinh x/2}\right)^2\right].
\end{split}
\label{eq:td-fep-stats}
\end{equation}
As one could have imagined, fluctuations exhibit a term growing linearly with $P$, 
which in the context of re-weighted sampling has particularly disruptive consequences.

The procedure for the SC-FEP estimator~\eqref{eq:sc-fep} 
is analogous. One evaluates the expectation value of the difference Hamiltonian
\begin{equation}
\begin{split}
\avg{\hY}_m&= \frac{(\alpha^{-1}-1) m\beta}{2P}\sum_{k>0} \omega^2\avg{\tilde{q}_k^2}_m=\\
&=\frac{\alpha^{-1}-1}{2} \sum_{k>0} \frac{1}{1+\omega_k^2/\omega^2} \underset{P\rightarrow\infty}{=} \\
&=\frac{\alpha^{-1}-1}{2} \left(\frac{x}{2}\coth\frac{x}{2}-1\right),
\end{split}
\end{equation}
which does not contain an explicit $P$ dependence in the asymptotic limit. This is also true of 
the fluctuations, that evaluate to
\begin{equation}
\begin{split}
\sigma^2(\hY)&=\left[\frac{(\alpha^{-1}-1) m\beta}{2P}\right]^2 2 \sum_{k>0} \left[\omega^2\avg{\tilde{q}_k^2}_m\right]^2=\\
&=\frac{(\alpha^{-1}-1)^2}{2} \sum_{k>0} \frac{1}{\left(1+\omega_k^2/\omega^2\right)^2}\underset{P\rightarrow\infty}{=} \\
&=\frac{(\alpha^{-1}-1)^2}{2} \left[\frac{x}{4}\coth \frac{x}{2} + \frac{1}{2}\left(\frac{x/2}{\sinh x/2}\right)^2-1\right].
\end{split}
\label{eq:sc-fep-stats}
\end{equation}
Eqs.~\eqref{eq:td-fep-stats} and~\eqref{eq:sc-fep-stats} capture the essential
statistical properties of the TD-FEP and SC-FEP estimators, and can be profitably used
to assess the range of applicability of the two methods by computing analytically the 
order of magnitude of the statistical errors one may expect for a system with known normal mode
frequencies.

\end{document}